\documentstyle[sprocl]{article}

\input{psfig}

\def\Journal#1#2#3#4{{#1} {\bf #2}, #3 (#4)}

\def\NPB{{\em Nucl. Phys.} B}
\def\PLB{{\em Phys. Lett.}  B}
\def\PRL{\em Phys. Rev. Lett.}
\def\PRD{{\em Phys. Rev.} D}

\def\st{\scriptstyle}

\def\be{\begin{equation}}
\def\ee{\end{equation}}
\def\bea{\begin{eqnarray}}
\def\eea{\end{eqnarray}}
\def\r2{\sqrt 2}
\def\sw2{\sin^2\theta_W}
\def\v#1{v_#1}

\def\c2b{\cos 2\beta}

\def\st{\tilde t}
\def\m#1{{\tilde m}_#1}
\def\mH{m_H}

\def\vW2{v_W^2}

\begin{document}

\title{Baryogenesis from top squark transport}

\author{Mayumi Aoki\footnote{Research Fellow of the Japan Society 
for the Promotion of Science.}
}

\address{
Graduate School of Humanities and Sciences \\
 Ochanomizu University  \\
Otsuka 2-1-1, Bunkyo-ku, Tokyo 112, Japan  \\
E-mail: amayumi@fs.cc.ocha.ac.jp} 

\author{Noriyuki Oshimo}

\address{
Institute for Cosmic Ray Research \\
 University of Tokyo  \\
Midori-cho 3-2-1, Tanashi, Tokyo 188, Japan  \\
E-mail: oshimo@icrr.u-tokyo.ac.jp}

%%%%%%%%%%%%%%%%%%%%%%%%%%%%%%%%%%%%%%%%%%%%%%%%%%%%%%%%%%%%%%
% You may repeat \author \address as often as necessary      %
%%%%%%%%%%%%%%%%%%%%%%%%%%%%%%%%%%%%%%%%%%%%%%%%%%%%%%%%%%%%%%

\maketitle\abstracts{
Baryon asymmetry of the universe is generated   
in the supersymmetric standard model (SSM)   
through the charge transport mechanism mediated by top squarks.  
Necessary $CP$ violation originates from a complex phase contained 
in mass-squared matrices of squarks and sleptons.  
This complex phase also gives contributions to the electric dipole moments
(EDMs) of the neutron and the electron.
The amount of induced baryon asymmetry is consistent with its 
observed value within reasonable ranges of SSM parameters.
The EDMs are predicted to be slightly smaller than their experimental
upper limits.
}

     Baryon asymmetry of the universe may be produced at the electroweak 
phase transition \cite{ewbrev}, 
since all the necessary conditions for baryogenesis seem to be satisfied 
within the framework of electroweak theories.  
However, 
$CP$ violation in the standard model (SM) arising from 
the Kobayashi-Maskawa phase is not large enough  
to produce a sufficient amount of asymmetry.
An extension of the SM is necessary 
to describe baryogenesis.

     In this note we discuss baryogenesis 
at the electroweak phase transition, assuming  
the supersymmetric standard model (SSM) based on $N$ = 1 supergravity 
coupled to grand unified theories.  
This model, which is one of the most plausible extensions of the SM, 
has new sources of $CP$ violation needed for baryogenesis.
It is shown \cite{aoki2} that a sufficiently large amount of baryon asymmetry
is produced through the charge transport mechanism \cite{ctm} mediated
by top squarks.
A complex phase in the mass-squared matrix of top squarks 
is the origin of $CP$ violation.  
This $CP$-violating phase also
contributes to the electric dipole moments (EDMs) of
the neutron and the electron. 
The EDMs are predicted to have fairly large values 
around their experimental upper limits.  

     The SSM has several complex parameters in addition to  
the Yukawa coupling constants for quarks and leptons.  
By redefining the phases of particle fields, without loss of generality, 
we can take as the physical complex parameters 
a Higgsino mass parameter $\mH$ for the bilinear term
of Higgs superfields in the
superpotential and several dimensionless coupling
constants $A_f$ for the trilinear terms of 
scalar fields which break supersymmetry softly.
We express these parameters as
$\mH = |\mH|\exp(i\theta)$ and $A_f = A = |A|\exp(i\alpha)$.  
Since $A_f$ are considered to have the same value of
order unity at the grand unification scale,
their differences at the electroweak scale are
small and thus can be neglected.
The phase $\theta$ is contained in 
the chargino and neutralino mass matrices and 
the squark and slepton mass-squared matrices.
The phase $\alpha$ is contained only in the latter.

     The new sources of $CP$ violation in the SSM give
contributions to the EDMs of the neutron and the electron
at the one-loop level through diagrams in which 
charginos, neutralinos or gluinos propagate together with
squarks or sleptons.
If the $CP$-violating phase $\theta$ is of order unity, 
these EDMs receive contributions dominantly from the chargino 
diagrams.  
From experimental constraints on the EDMs,
the masses of squarks and sleptons are predicted
to be larger than 1 TeV \cite{edm}.  
In this case, top squarks are out of thermal equilibrium at the 
electroweak phase transition, 
and thus cannot mediate the charge transport mechanism. 
We assume that the phase $\theta$ is sufficiently smaller
than unity, making the chargino contributions to the EDMs negligible.  
Then, sizable new $CP$-violating phenomena 
can only be induced by the phase $\alpha$.  
The neutron and electron EDMs receive dominant
contributions respectively from the gluino and neutralino diagrams, 
leading to relaxed constraints on the masses of $R$-odd 
particles \cite{edm}.  
A recent analysis shows \cite{aoki3} that 
even if the phase $\alpha$ is of order unity,  
squarks and sleptons 
are allowed to have masses of order 100 GeV under the conditions 
$\mH\sim 100$ GeV and $\m2>500$ GeV, $\m2$ being the SU(2) 
gaugino mass.       

     For the charge transport mechanism to work, 
the electroweak phase transition has to be strongly first order.  
Accordingly, the Higgs potential is constrained, leading to 
an upper limit for the mass of a Higgs boson.  
Although this limit in the SM is not compatible with experimental 
results, that in the SSM has not yet been ruled out.  
Furthermore, if a top squark is light, quantum corrections 
make it easy for the Higgs potential to yield a phase transition 
of strongly first order.  We thus assume an electroweak 
phase transition of strongly first order in the following 
discussions.   

     At the electroweak phase transition,  
bubbles of the broken phase nucleate in the
SU(2)$\times$ U(1) symmetric phase.  
Neglecting generation mixings, the left-handed top squark 
$\st_L$ and the right-handed top squark $\st_R$ are in
mass eigenstates in the symmetric phase,
where the vacuum expectation values (vevs) $\v1$ and $\v2$ 
of Higgs bosons vanish.
In the broken phase and the bubble wall,
they are mixed to form mass eigenstates,
owing to non-vanishing vevs.
At the bubble wall, $\st_L$ coming from the symmetric phase
can be reflected to become $\st_R$, and vice verse.
The mass eigenstates of top squarks $\st_1$ and 
$\st_2$ in the broken phase
can be transmitted to the symmetric phase and become $\st_L$ or $\st_R$.
In these processes, the complex phase $\alpha$ in the mass squared-matrix 
of top squarks induces $CP$ violation, causing differences in
reflection and transmission rates between $CP$ conjugate processes.  
Although other squarks and sleptons also have $CP$-violating rates for 
the transitions,   
the resultant asymmetries are neglected 
because of the small Yukawa coupling constants of corresponding quarks 
and leptons.  

     The transition rates at the wall are obtained 
by solving Klein-Gordon equations.   
For the profile of the bubble wall, we assume a 
hyperbolic-tangent dependence of $\sqrt{\v1^2+\v2^2}$ 
on the space coordinate, 
changing from 0 to the value of the broken phase.  
The ratio $\v2/\v1$
must also vary in the wall for $CP$ violation to occur.  
We assume that it changes linearly by $\Delta(\v2/\v1)\sim 1-10$.  
The wall width $\delta_W$ has been estimated in the SM as 
$\delta_W\sim 10/T$ \cite{turok}, where $T$ stands 
for the temperature of the electroweak phase transition.  
This width  generally becomes thinner if the phase transition 
is strongly first order.   

     The induced $CP$ asymmetries of  
the reflection and transmission rates cause  
a net flux of hypercharge emitted from the bubble wall 
into the symmetric phase.  
In the symmetric phase,  
the chemical potential $\mu_B$ of the baryon number is related to  
the hypercharge density $\rho_Y$ through equilibrium conditions.  
Taking the densities for the total baryon number,
baryon number of the
third generation, and lepton number to approximately vanish,
we obtain the relation $\mu_B=-2\rho_Y/9T^2$.  
A net hypercharge density in front of the wall makes $\mu_B$ 
non-vanishing, which favors a non-vanishing value of the baryon number.

     The baryon number changes   
through an electroweak anomaly.  This rate  
is not small in the symmetric phase, contrary to its 
negligible smallness  in the broken phase. 
Assuming detailed balance for the transitions among
the states of different baryon numbers, 
the rate equation of the baryon number density $\rho_B$ is given by 
$d\rho_B/dt=-(\Gamma/T)\mu_B$, $\Gamma=3\kappa(\alpha_WT)^4$, where 
$\Gamma$ denotes the rate for the transition and 
$\kappa$ is estimated as $0.1-1$ \cite{ambjorn}.
Solving this equation, the ratio of the baryon number density to the entropy 
density $s$ is given by
$\rho_B/s=15\kappa\alpha_W^4F_Y\tau_T/ \pi^2g_*v_W T^2$, 
where $g_*$ represents the relativistic degree of freedom
for the particles in thermal equilibrium, 
$\tau_T$ is the time which the carriers of the hypercharge
flux spend in the symmetric phase before being captured
by the wall, $v_W$ is the velocity of the wall,
and $F_Y$ is the hypercharge flux emitted from the wall.

     We now calculate the baryon asymmetry  
for $\theta \ll 1$ and $\alpha \sim 1$.
The mass parameters for squarks and sleptons  
are taken to be of order 100 GeV.
The Higgsino mass parameter $|\mH|$ is smaller than or around 
these mass parameters to have a neutralino lighter than 
the squarks and sleptons, as required on cosmological grounds.  
For definiteness, we take $g_*=214.75$,
where all the particles excluding the gluinos are
taken into account.
The time $\tau_T$
may be approximated by the mean free time of the carriers 
of the hypercharge flux.
A rough estimate  gives $\tau_T\sim 10/T$ for quarks \cite{joyce},
which is considered to be applicable to squarks.
The wall velocity is estimated as $v_W=0.1-1$ \cite{turok}.
The temperature $T$ is of order 100 GeV.  

     The obtained ratio of baryon number to entropy is 
$\rho_B/s=(1-5)\times 10^{-11}$ for $\kappa\sim 1$,
which is compatible with the observed value.  
The sign of $\rho_B$ depends on the phase $\alpha$ and on 
whether the ratio $\v2/\v1$ increases or decreases in the wall.   
If the phase $\alpha$ becomes of order 0.1, it is difficult to produce
a sufficient degree of asymmetry.  
The value of $\kappa$ should also be of order unity.  

     The $CP$-violating phase $\alpha$ induces  
the EDMs of the neutron and the electron, respectively, through 
the gluino and the neutralino diagrams.
The gluino mass is determined by the SU(2) gaugino mass $\m2$.
The neutralino masses are determined by $\m2$, $\mH$, and the 
ratio $\v2/\v1$.  
The masses of the squarks and 
sleptons of the first generation are roughly 
the same as those of the third generation.  
Among these parameters on which the EDMs depend, the 
SU(2) gaugino mass $\m2$ is not directly related to the baryon 
asymmetry.  
In the parameter ranges where the baryon asymmetry can be explained,
the magnitude of the neutron EDM is $10^{-25}-10^{-26}e$cm 
for $\m2=500-1000$ GeV.  
This is consistent with the experimental constraint, though 
not much smaller than the experimental upper limit.  
For $\m2<500$ GeV,   
the EDM becomes larger than this limit.   
The electron EDM is predicted to be $10^{-26}-10^{-27}e$cm  
in the same ranges of parameters, which is also slightly smaller
than its experimental upper limit.  

     In the SSM the charge transport mechanism for baryogenesis 
could also be mediated by charginos or neutralinos.  
For $\theta\sim 1$,
although squarks and sleptons have to be rather heavy, 
the charginos and neutralinos 
can have masses of order 100 GeV.  
A sufficiently large amount of baryon number is actually produced  
by the charginos \cite{aoki}.
In this case, the neutron and electron EDMs are predicted to be 
$10^{-25}-10^{-26} e$cm and 
$10^{-26}-10^{-27} e$cm, respectively, 
for the squark and slepton masses of $1-10$ TeV. 

     In summary, we have studied the possibility of 
baryogenesis in the SSM.    
The baryon asymmetry can be produced sufficiently 
through the charge transport mechanism mediated by top squarks  
with masses of order 100 GeV,
if one $CP$-violating phase $\alpha$ is not suppressed
while another phase $\theta$ kept small. 
The EDMs of the neutron and the electron
are predicted to have magnitudes   
not so small compared to their present
experimental upper limits. 

\section*{Acknowledgments}

     We thank J.W.F. Valle and the staff of the organizing committee 
for their kind hospitality.
The work of M.A. is supported in part by the Grant-in-Aid
for Scientific Research from the Ministry of Education, Science
and Culture of Japan.
This work is supported in part by the Grant-in-Aid for Scientific
Research (No. 08044089, No. 08640357, No. 09246211) 
from the Ministry of Education, Science and
Culture, Japan.

\newpage 
\pagestyle{empty}

%\begin{figure}
\psfig{file=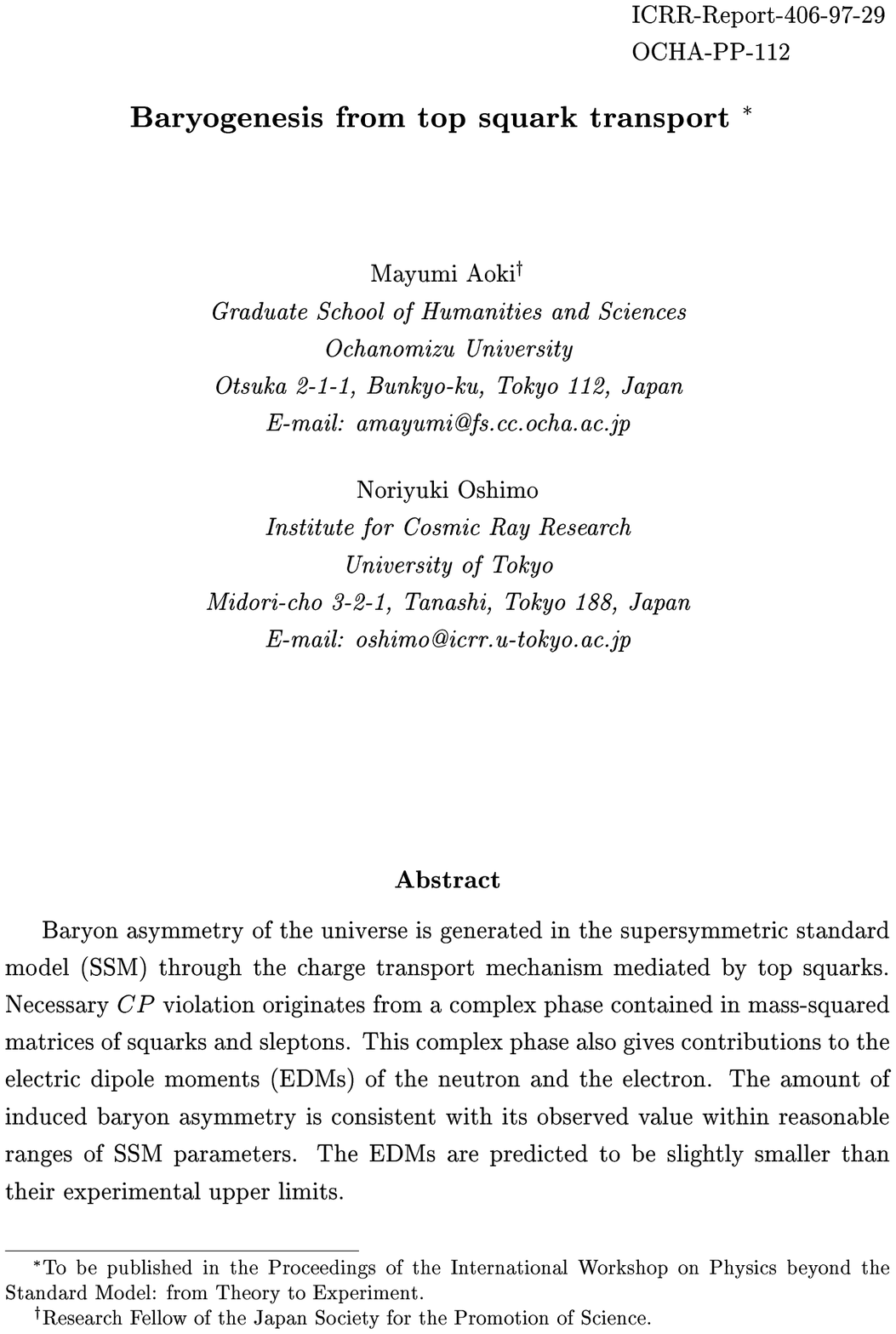}
%\end{figure}

\end{document}